\documentclass{mn2e-gfl}

\usepackage{epsfig}
\usepackage{natbib}

\newcommand{\ltaraw}{$\; \buildrel < \over \sim \;$}
\newcommand{\lta}{\lower.5ex\hbox{\ltaraw}}
\newcommand{\gtaraw}{$\; \buildrel > \over \sim \;$}
\newcommand{\gta}{\lower.5ex\hbox{\gtaraw}}

\loadboldmathitalic
\title[The Cetus Dwarf Galaxy]{
Inside the whale: the
structure and dynamics of the isolated Cetus dwarf spheroidal}
\author[Lewis et al.]{
G.F. Lewis$^{1}$, 
R.A. Ibata$^{2}$,
S.C. Chapman$^{3}$,
A. McConnachie$^{4}$, 
 \newauthor 
M.J. Irwin$^{5}$,
E. Tolstoy$^{6}$ \&
N.R. Tanvir$^{7}$\\
$^{1}${Institute of Astronomy, School of Physics, A28, 
University of Sydney, NSW 2006, Australia:
{\tt gfl@physics.usyd.edu.au}}\\
$^{2}${Observatoire de Strasbourg, 11, rue de l'Université, 
Strasbourg F-67000, France}\\
$^{3}${California Institute of Technology, Pasadena, CA 91125}\\
$^{4}${University of Victoria, Dept. of Physics \& Astronomy, 
Victoria, BC VP8 1A1, Canada}\\ 
$^{5}${Institute of Astronomy, Madingley Road, Cambridge CB3 0HA, UK}\\
$^{6}${Kapteyn Institute, University of Groningen, Postbus 800, 
9700 AV Groningen, Netherlands.}\\ 
$^{7}${Physical Science, University of Hertfordshire, Hatfield, 
AL10 9AB, UK}
}
\date{\today}
\begin{document}
\maketitle
\label{firstpage}
\begin{abstract}
  This paper  presents a study  of the Cetus  dwarf, an isolated  dwarf galaxy
  within the Local Group.  A matched-filter analysis of the INT/WFC imaging of
  this system reveals no evidence for significant tidal debris that could have
  been torn  from the galaxy, bolstering  the hypothesis that  Cetus has never
  significantly  interacted with either  the Milky  Way or  M31. Additionally,
  Keck/Deimos spectroscopic  observations identify  this galaxy as  a distinct
  kinematic population  possessing a  systematic velocity of  $-87\pm2{\rm km\
    s^{-1}}$ and with a velocity dispersion of $17\pm2{\rm km\ s^{-1}}$; while
  tentative,  these  data  also   suggest  that  Cetus  possesses  a  moderate
  rotational velocity of $\sim8{\rm km\ s^{-1}}$.  The population is confirmed
  to be  relatively metal-poor,  consistent with ${\rm  [Fe/H]\sim-1.9}$, and,
  assuming virial equilibrium, implies that the Cetus dwarf galaxy possesses a
  $M/L\sim70$. It  appears, therefore, that  Cetus may represent  a primordial
  dwarf  galaxy, retaining  the kinematic  and structural  properties  lost by
  other  members  of  the  dwarf  population  of  the  Local  Group  in  their
  interactions with the  large galaxies. An analysis of  Cetus's orbit through
  the Local Group indicates that it is at apocentre; taken in conjunction with
  the general dwarf  population, this shows the mass of the  Local Group to be
  $\gta2\times10^{12}M_\odot$.
\end{abstract}

\def\newblock{\hskip .11em plus .33em minus .07em}
\section{Introduction}
Within the  Cold Dark Matter (CDM)  paradigm, galaxies grow  over time through
the continued accretion of smaller systems~\citep{1978MNRAS.183..341W}.  Dwarf
galaxies,  therefore,  being at  the  bottom  of  this hierarchical  formation
picture  (or food  chain),  are  amongst the  most  `fundamental' of  galactic
building blocks.   While recent metallicity observations of  Local Group dwarf
galaxies has suggested  that the Milky Way could not  have formed from systems
resembling the  current dwarf population  \citep{2004AJ....128.1177V}, current
accretion          events          in          the          Milky          Way
\citep{1994Natur.370..194I,2004MNRAS.348...12M}     and    Andromeda    Galaxy
\citep{2001Natur.412...49I}  illustrate that the  galaxy formation  process is
still ongoing, albeit at a gentle pace.

Dwarf galaxies  remain the  dominant population of  galaxies in  the Universe;
while the Local Group contains only two dominant, large galaxies (Milky Way \&
M31),    the    population   of    dwarf    galaxies    numbers   more    than
40~\citep{1998ARA&A..36..435M}. The  advent of deep, panoramic  surveys of the
Local  Group has  recently revealed  several more  examples of  these galaxies
\citep{2004ApJ...612L.121Z,2005ApJ...626L..85W,2005AJ....129.2692W,2006ApJ...642L.137B,2006ApJ...643L.103Z,2006ApJ...647L.111B,2006ApJ...650L..41Z},
but still the  apparent conflict between the expected  and observed population
of  dwarf  galaxies  remains  \citep{1999ApJ...522...82K,1999ApJ...524L..19M}.
The theoretical  and observational  picture can be  brought into  agreement if
star  formation in  low  mass systems  is  suppressed via  such mechanisms  as
photoionization  in  the  early   universe,  leaving  only  a  meager  stellar
population to reveal their location.  Such models, however, predict that dwarf
galaxies   should   be   enveloped   in   massive   halos   of   dark   matter
\citep{2002MNRAS.335L..84S};  such haloes  would  be revealed  in the  stellar
kinematics of the dwarf at large radii.

Dwarf spheroidal galaxies (generally gas-poor, pressure supported systems) are
preferentially found  as satellites  to the Milky  Way and M31,  whereas dwarf
irregular galaxies  (generally gas rich, rotating  systems) are preferentially
found  as  isolated  systems.   This position-morphology  relation  was  first
highlighted by  \citet{1974Natur.250..309E} and suggests  that the environment
of  a  dwarf  galaxy  plays  a  fundamental role  in  driving  its  evolution.
\citet{2001ApJ...559..754M,2001ApJ...547L.123M,2006MNRAS.369.1021M}     suggest
that tidal effects and ram pressure stripping of dwarf galaxies in the halo of
large galaxies may be sufficient  to turn a rotationally supported system into
a  pressure  supported system.   Isolated  dwarf  spheroidals are,  therefore,
particularly interesting  as they would have  been immune for  the shaking and
stirring that has  occurred to most dwarf systems and  hence may represent the
dwarf  galaxies in  their most  pristine form.   However, such  isolated dwarf
spheroidal  galaxies are  rare within  the Local  Group, with  two  dwarfs not
currently  a   satellite  of  either  the   Milky  Way  and   M31;  Cetus  and
Tucana\footnote{The transition  type dwarf spheroidal/irregular,  DDO210, lies
  $\sim1$ Mpc from both M31 and the Milky Way, and may be interacting with the
  Local  Group  for  the  first time~\citep{alaninprep2}.}.   Determining  the
dynamical properties and  orbital history of such galaxies  can help determine
the role of interactions in governing a dwarf galaxy's morphology.

Discovered   by   \citet{1999AJ....118.2767W}  in   an   eye-ball  survey   of
photographic plates,  the Cetus dwarf galaxy is  seen to lie at  a distance of
$755\pm23$kpc  \citep[][ assuming $E(B-V)=0.029$]{2005MNRAS.356..979M};  it is
almost as  remote from M31,  with a separation  of 680kpc.  Hence,  unlike the
vast majority of  dwarf systems within the Local Group,  Cetus does not appear
to be part of  the entourage of either the Milky Way  or Andromeda Galaxy, and
apparently  represents  an isolated  member  of  the  local dwarf  population.
Furthermore, an analysis  of the stellar distribution in  this dwarf reveal it
to  possess a half-light  radius of  more than  300pc, and  a tidal  radius of
6.6kpc\footnote{As noted in \citet{alaninprep},  the tidal radius for Cetus is
  very  poorly  constrained,  with  no  significant turnover  in  the  surface
  brightness  profile  at  the  largest  radii  probed.   If  it  has  had  no
  interaction, the tidal radius may not represent a physical limit to Cetus.},
making  it  the  largest  radial  extent dwarf  within  the  local  population
\citep{alaninprep}.  Such a large physical  extent suggests that Cetus has not
undergone a significant interaction with  other members of the Local Group and
hence  provides an  ideal  testbed for  kinematic  analysis.  To  this end,  a
reanalysis  of the  INT/WFC data  obtained by  \citet{2005MNRAS.356..979M} was
undertaken to search for any evidence  of tidal disruption due to any previous
interactions.  This was coupled with a spectroscopic survey of the Cetus dwarf
galaxy which  was undertaken with  Deimos on the  Keck 10m telescope  with the
goal of determining its kinematic  and dark matter properties.  The details of
the  observations   are  presented  in   Section~\ref{observations},  while  a
discussion of the results appears in Section~\ref{results}.  This paper closes
with Section~\ref{conclusions} which presents the implications and conclusions
of this study.

\begin{figure}
\centerline{ \psfig{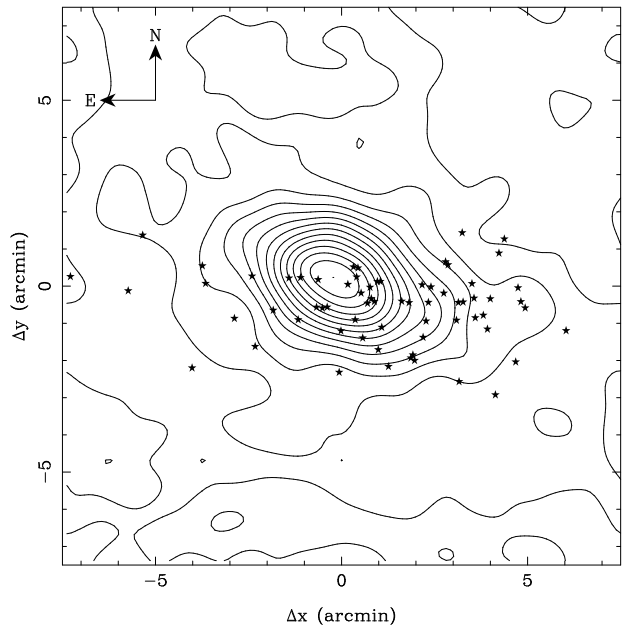}}
\caption[]{The contours in this figure denote the stellar density in the Cetus
  dwarf   galaxy    as   derived   from   INT/WFC    imaging,   centred   upon
  $\alpha=00^h26^m11^s\              \delta=-11^o02'40''              (J2000)$
  \citep[see][]{alaninprep}. The stars symbols  represent the locations of the
  stars for which form the basis of this current study.
  \label{fig1}}
\end{figure}

\section{Observations}\label{observations}

\subsection{Photometric}\label{photometric}
\begin{figure*}
\centerline{ \psfig{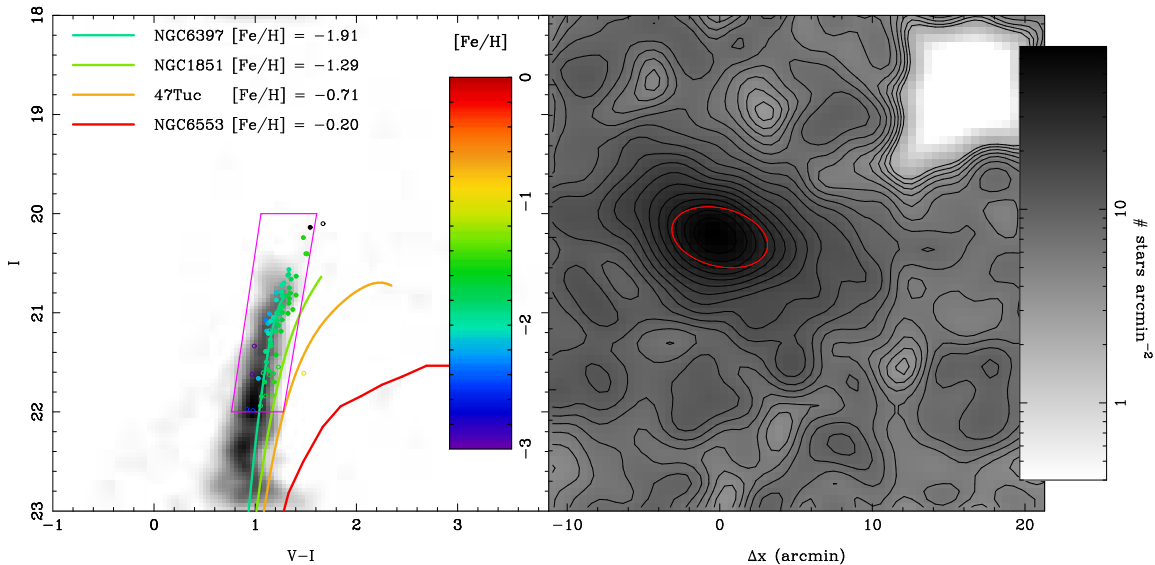}}
\caption[]{The left-hand
panel presents the match filter weight map, constructed by 
dividing the CMD within the
central 3  arcmins of  the Cetus  dwarf by  the CMD  from the
region beyond 10 arcmins of the dwarf; this enhances the CMD signature
of  the  dwarf.  The  isochrones  from  several  fiducial systems  are
overplotted  for  comparison. The  closed  box  denotes the  selection
region as  described in Section~\ref{photometric},  while the coloured
circles  represent the observed  stars, with  solid circles  are those
within 30  ${\rm km\ s^{-1}}$ of  the systemic velocity  of Cetus. The
colour-coding represents the spectroscopically determined metallicity,
as  described in  Section~\ref{colour} and  defined by  the  key.  The
right-hand  panel presents  the logarithm of the stellar density as 
identified by the matched filtering analysis. The 
peak value is $\sim70$ stars per square arcmin, with each contour 
corresponding to 0.11dex; this is also represented in the greyscale as
given in the key. The  red  ellipse corresponds to the stellar 
peak seen in Figure~\ref{fig1}, whereas the
square region in the upper  right hand corner represent the region not
covered by the INT/WFC field.
\label{fig2}}
\end{figure*}

The photometric observations were obtained  with the INT/WFC on the 2.5m Isaac
Newton   Telescope   on   La   Palma   and  were   originally   presented   by
\citet{2005MNRAS.356..979M}.   As noted  in this  previous work,  the V  and I
colour magnitude diagram of the Cetus  dwarf clearly reveals the presence of a
red-giant  branch below  I=20.5. Furthermore,  the  presence of  the dwarf  is
clearly revealed in stellar density, as apparent in Figure~\ref{fig1}.  Hence,
this photometric  data was used  to select targets for  spectroscopic followup
with the Keck telescope. The stars were chosen to lie within $\sim0.15$mags of
the  Red  Giant Branch  (RGB)  between  $22.0<I<20.5$;  the selection  box  is
overlaid  on  the  CMD   presented  in  Figure~\ref{fig2}  and  the  spatially
distribution  of stars appears  in Figure~\ref{fig1}.   The selection  box was
extended to $I=20$ to include candidate AGB stars into the sample.

\subsection{Spectroscopic}\label{spectroscopic}
Observations     were    undertaken     with    the     DEIMOS    spectrograph
\citep{2003SPIE.4841.1657F} on the  10m W.  M. Keck-II telescope  in Hawaii on
the night of  the $1^{\rm st}$ October, 2005. The  1200l/mm grating was blazed
to  800nm and,  coupled  with  the OG550  order-blocking  filter, provided  an
effective  spectral range  from 700nm  to 900nm,  encompassing the  calcium II
triplet (CaT); it  is this series of lines that  provide kinematic and overall
abundance ([Fe/H]) measures.

Custom made slit masks were used to target stars in the Cetus dwarf galaxy; as
noted previously, the  astrometry was drawn from the  INT/WFC survey presented
in \citet{2005MNRAS.356..979M}, and the target  field is shown overlain on the
stellar  density map  of  the  dwarf in  Figure~\ref{fig1}.   The fields  were
flat-fielded and sky-subtracted as part  of the standard Deimos data reduction
pipeline,  while spectra  were extracted  using a  dedicated  software package
\citep[e.g.][]{2005ApJ...634..287I}.  Finally,  velocities were calculated via
cross-correlation  of  the   spectra  with  zero-velocity  stellar  templates,
utilizing the CaT, centred at  $\sim8550{\rm \AA}$.  This resulted in 70 stars
with a  velocity error  less than  20${\rm km\ s^{-1}}$  with a  mean (median)
velocity error of 8.8 (7.7) ${\rm km\ s^{-1}}$; the data from these stars are 
presented in Table~1 and 2.

\begin{figure}
\centerline{ \psfig{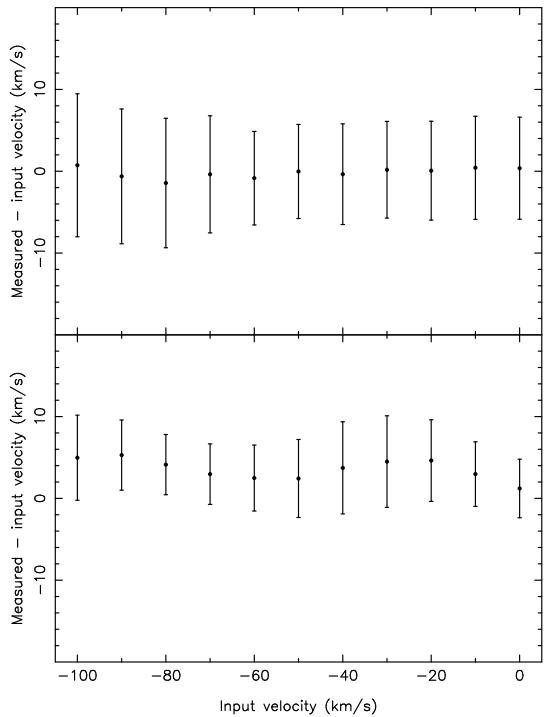}}
\caption[]{The systematic error in velocity as determined from the 
analysis of synthetic spectra, using sky residuals as determined from 
the data. The error bars represent the uncertainty in the velocity 
determination. The top panel uses Poisson sampling of the median sky,
whereas the the lower panel uses the median skies from separate CCDs.
The overall results show  no substantial velocity systematics in 
the range of $-100\rightarrow0 {\rm km\ s^{-1}}$ are typically 
$\sim 1 {\rm km\ s^{-1}}$ in the upper panel, increasing to 
$\sim 4-5 {\rm  km\ s^{-1}}$ in the lower panel.
\label{fig3}}
\end{figure}

While the  standard velocity  errors are well  understood, it is  important to
estimate  any potential  systematic  uncertainties that  may  result from  the
reduction procedure.  To  this end, fake stellar spectra in  the region of the
CaT were  generated, using sky  residuals determined from Poisson  sampling of
the median observed sky spectrum.  The resultant synthetic spectra possessed a
S/N$\sim5$  and were  analyzed in  the  same fashion  as the  real data.   The
extracted velocity is compared to  the input velocity in Figure~\ref{fig3}; in
this  figure,  the top  panel  presents  the  results from  synthetic  spectra
generated  using the  overall median  sky  spectrum, whereas  the lower  panel
utilized median  sky spectra  from separate CCDs  and hence represents  a more
extreme estimate of the sky residuals. It  can be seen that in the region from
$-100\rightarrow 0 {\rm km\ s^{-1}}$, the upper panel reveals small systematic
uncertainties $(\sim  1 {\rm km\  s^{-1}})$, whereas the lower  panel displays
more significant  uncertainties of  $(\sim 4-5 {\rm  km\ s^{-1}})$. It  can be
concluded that  in this region,  the systematic uncertainties in  the velocity
determination are roughly half the value of the mean standard error.

\section{Results}\label{results}

\subsection{Photometric Search for Extra-Tidal Stars}
Given  its remote  location, Cetus  appears to  have lived  a  rather isolated
existence, free from the large galactic tides that have influenced other dwarf
galaxies. However, given the uncertainty of its orbit, its detailed historical
motion through the Local Group is unknown and there is the possibility that it
underwent interactions  with other  members in the  past.  The result  of such
interactions would  depend upon the  specifics of the collisions,  although it
would be  expected that if  it interacted with  a substantial system  then the
tidal  disruption would result  in the  formation of  tidal tails  which could
still accompany the dwarf today \citep[e.g.][]{2002AJ....124..127J}.

Tidal debris, however, typically presents a very low surface brightness and is
not  apparent  in  maps  of  stellar density.   However,  various  statistical
techniques exist that allow the  extraction of low surface brightness features
by identifying the sequence of the dwarf within an overall background CMD.  To
this          end,         a          matched          filter         analysis
\citep{1999ApJ...517...78K,2002AJ....124..349R}  of the  INT/WFC  data of  the
Cetus dwarf was  undertaken; for this, a background  CMD was constructed using
stars located more  than 10 arcmins from the centre of  the Cetus dwarf, where
the contribution  of stars from Cetus  is very small relative  to the Galactic
foreground, while the CMD of the  dwarf was obtained from the inner 3 arcmins;
this is presented in the left-hand panel of Figure~\ref{fig2}, overlaid with a
series of fiducial  sequences, confirming the low metallicity  of this system.
The right-hand panel  of this figure presents the logarithm  of the density of
stars associated with the dwarf  as identified by the matched filter analysis,
smoothed with a Gaussian kernel of 2 arcmins; note that the large blank region
in the upper right-hand  part of this figure is an artifact  of the CCD layout
of the  INT/WFC.  The red ellipse, with  semi-major and minor axes  of 2.0 and
3.3  arcmins, encloses  the  peak  of the  stellar  distribution presented  in
Figure~\ref{fig1}. The greyscale  (and contours) correspond to the  key on the
right-hand  side of  this panel  and,  as presented  in \citet{alaninprep},  a
stellar density  of $\sim1$ star  per ${\rm arcmin^2}$ corresponds  to $\sim30
{\rm mags\  arcmin^{-2}}$.  The $3\sigma$  noise limit in the  stellar density
map  is  $\sim\pm  2$ stars  per  ${\rm  arcmin^2}$,  resulting in  a  surface
brightness limit of $\sim29 {\rm  mags\ arcmin^{-2}}$; hence, any extension of
the    dwarf     leading    to     tidal    tails    should     be    apparent
\citep[e.g.][]{2002MNRAS.336..119M}.    An  inspection   of  Figure~\ref{fig2}
clearly  reveals the  presence of  the  Cetus dwarf  itself and,  as noted  in
\citet{alaninprep}, and  apparent in Figure~\ref{fig1}, the  dwarf possesses a
rather  elliptical   morphology;  while  such  a   non-spherical  shape  could
potentially be interpreted  as a signature of previous  interactions, the lack
of other  evidence of interactions  suggests this is  not the case and  we can
conclude Cetus  was formed with  this morphology.  Furthermore,  inspection of
the stellar density in the vicinity of  the main body of the dwarf uncovers no
coherent, stream-like structures that could  be interpreted as tidal tails and
debris  \citep[c.f.    the  magnificent  tidal   stream  of  Pal  5   as  seen
by][]{2002AJ....124..349R,2006ApJ...641L..37G};  while there  appears to  be a
slight signature to the lower right of  the dwarf itself, it is clear from the
scale of the noise over the image, this could not be considered significant.

\subsection{Kinematics}\label{kinematics}

\begin{figure}
\centerline{ \psfig{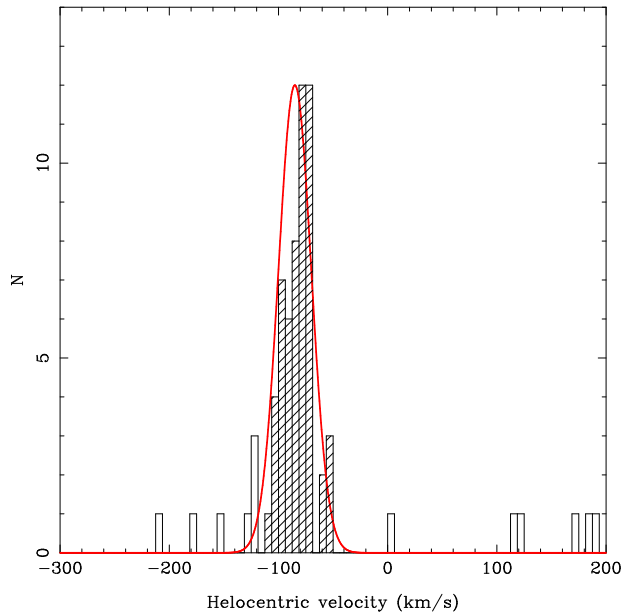}}
\caption[]{The  distribution of  stellar velocities  obtained  in this
survey;  a clear over-density,  denoting the  systemic velocity  of the
Cetus dwarf  is clearly visible, with a  negligible contamination from
the Galaxy. The overplotted solid line  is a Gaussian fit to the data,
assuming a systemic  velocity of -85${\rm km\ s^{-1}}$  and a width of
15${\rm  km\  s^{-1}}$. The  dashed  histogram corresponds  highlights
those stars within $\pm30{\rm km\ s^{-1}}$ of the systemic velocity.
\label{fig4}}
\end{figure}

Figure~\ref{fig4} presents the distribution of Heliocentric stellar velocities
obtained from the Keck/Deimos spectra, with a clear peak denoting the presence
of  the Cetus  dwarf  galaxy.  Overplotted  on  this figure  is  the best  fit
Gaussian,  with a  systemic velocity  of -85${\rm  km\ s^{-1}}$  and  width of
15${\rm km\ s^{-1}}$; note that  this is inconsistent with velocities of three
HI clumps (-311,  -268 \& -262${\rm km\ s^{-1}}$) in the  vicinity of Cetus as
measured by \citet{2006AJ....131.2913B}, and it must be concluded that they do
not represent Cetus's  gas content.  Following \citet{1986AJ.....92...72R}, we
assume the Cetus  dwarf can be represented as  a simple, spherically symmetric
stellar  system  of  central  surface brightness  $\Sigma_o$,  half-brightness
radius of $r_{hb}$ and velocity  dispersion $\sigma$ and can then estimate the
mass-to-light ratio in the system to be
\begin{equation}\label{masstolight}
\frac{M}{L} = \eta \frac{9}{2\pi G} \frac{\sigma^2}{\Sigma_o r_{hb}}
\end{equation}
where $\eta$ is a dimensionless  structure parameter which has a typical value
of $\sim$unity; this formula assumes that mass traces light within the dwarf.

\begin{figure}
\centerline{ \psfig{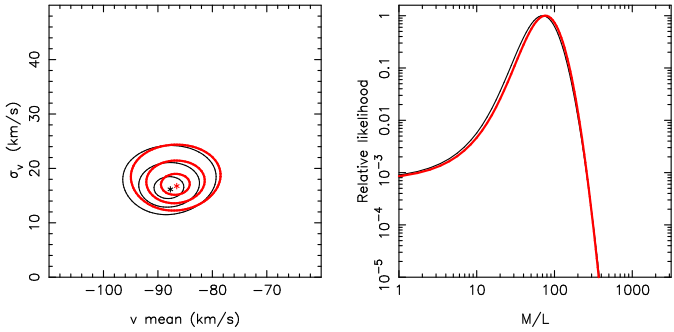}}
\caption[]{The left-hand panel presents  the likelihood contours (at $1\sigma,
  2\sigma$ and $3\sigma$) for the mean and dispersion of the radial velocities
  of  the Cetus  dwarf  populations.  The black  curves  represent the  entire
  sample, while  the red curves represent  stars within $5'$ of  the centre of
  the dwarf. The right-hand panel  presents the relative likelihood curves for
  the  M/L ratio  of the  dwarf  (given the  same selection  as the  left-hand
  panel); both peak strongly in the region of M/L$\sim70$.
\label{fig5}}
\end{figure}

Figure~\ref{fig5} presents  the results of  a likelihood calculation  for this
population of  RGB stars. Firstly,  the right-hand panel presents  contours at
$1\sigma, 2\sigma$ and $3\sigma$ of  the mean velocity and velocity dispersion
of   this  population,   consistent   with  the   fit   values  presented   in
Figure~\ref{fig4}. The dark line in  this figure corresponds to the entire RGB
sample, whereas the lighter line utilizes  only those stars within $5'$ of the
centre of  the dwarf; the results  of this analysis suggest  that the systemic
velocity of  Cetus is  $-87\pm2$ ${\rm km\  s^{-1}}$ with an  internal velocity
dispersion of $17\pm2$ ${\rm km\  s^{-1}}$.  The left-hand figure presents the
relatively likelihood for the mass to  light ratio of the Cetus dwarf, clearly
peaking at a $M/L\sim70$.

%
%

\subsection{Spectroscopic Metallicities}\label{colour}
The equivalent widths of the CaT can be used as a proxy measure of the stellar
metallicity.      As     with     our    previous     Keck/Deimos     studies,
\citep[e.g.][]{2005ApJ...634..287I},          the          approach         of
\citet{1997PASP..109..907R} is adopted, with;
\begin{equation}
\left[\frac{Fe}{H}\right] = -2.66 + 0.42\left[\sum Ca - 0.64(V_{HB} - V)\right]
\end{equation}
where
\begin{equation}
\sum Ca = 0.5 EW_{\lambda 8498} + EW_{\lambda 8542} + 0.6 EW_{\lambda 8662}
\end{equation}
and the $(V_{HB}  - V)$ provides a surface gravity  correction term; for Cetus
$V_{HB}    =   25.10$,    scaled   from    the   M31    value    employed   by
\citet{2005ApJ...634..287I}.  The  spectroscopically derived metallicities are
presented as  colour-coded points are  superimposed on the left-hand  panel of
Figure~\ref{fig2};  the  solid circles  in  this  figure  denote stars  within
30${\rm km\  s^{-1}}$ of the systemic  velocity, whereas the  open circles are
the remaining stars in the sample.  The errors on the metallicity measures are
$\sim0.2$.   The  stars  overlay  several  fiducial  tracks  of  systems  with
determined metallicities  (which are  colour-coded on the  same scheme  as the
stars).  A  comparison of  the stellar metallicities  and the  fiducial tracks
reveal that the two measures are consistent and that the Cetus dwarf galaxy is
dominated by an  old, metal poor population with  $[Fe/H]\sim-1.9$, as per the
findings of \citet{1999AJ....118.2767W}.

\subsection{A Hint of Rotation}\label{rotation}
\begin{figure}
\centerline{ \psfig{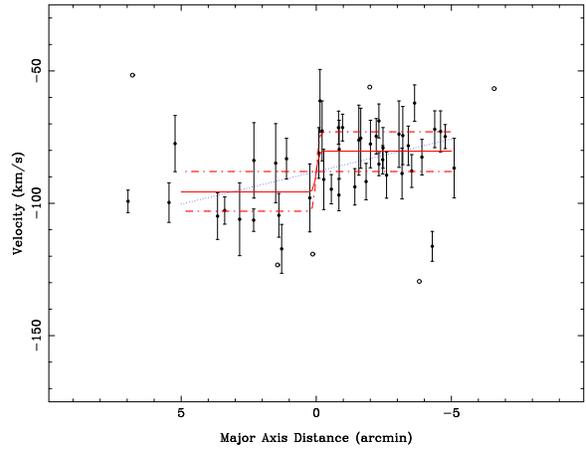}}
\caption[]{The spatial  distribution of stars along the  major axis of
the Cetus dwarf  galaxy. The solid points denote  stars within 30${\rm
km\ s^{-1}}$ of  the systemic velocity. The solid  line represents the
best  fit  rotational velocity  of  7.7${\rm  km\  s^{-1}}$, with  the
dot-dashed  line representing  fiducial rotation  velocities of  0 and
15${\rm km\ s^{-1}}$. The dotted line presents a linear model, representing
solid body rotation, with a best fit of $2.5\pm0.4{\rm km\ s^{-1} arcmin^{-1}}$.

\label{fig6}}
\end{figure}

Figure~\ref{fig6} presents  the spatial distribution of stars  along the major
axis,  with velocities  within 30${\rm  km\ s^{-1}}$  of the  systemic (filled
circles).  An eye-ball  examination of this figure reveals  that there appears
to be a velocity offset between those stars to the right and left of the dwarf
centre, with those  at positive distances possessing a  more negative velocity
than those  at negative distance.  Noting  the errors to the  data points, the
signature is somewhat tentative and  hence no rigorous kinematic model fitting
was  undertaken. As  an illustration,  a simple  constant  rotational velocity
profile was fitted to  the stars within 5 arcmins of the  centre of the dwarf,
with a resultant best fit rotational velocity of $7.7\pm1.2 {\rm km\ s^{-1}}$;
this rotation  curve is  overplotted as a  solid line in  Figure~\ref{fig6} as
well as two additional fiducial rotation curves of 0 and 15${\rm km\ s^{-1}}$.
The dotted line  represents a linear model, representing  solid body rotation,
with a best  fit of $2.5\pm0.4{\rm km\ s^{-1}  arcmin^{-1}}$.  In removing the
signature of the best fit  constant rotation velocity, the velocity dispersion
within the  inner 5 arcmins  falls to 14  ${\rm km\ s^{-1}}$ with  a 4$\sigma$
clipping),  and  to  8 ${\rm  km\  s^{-1}}$  when  applying a  more  stringent
2$\sigma$  clipping.  Hence,  if  real,  the rotation  of  Cetus represents  a
significant kinematic component in this system.

\subsection{Constraints on the Local Group}\label{localgroup}
As noted  previously, Cetus is one of  only two isolated dSph  galaxies in the
Local Group.  Hence, measurement of its heliocentric velocity provides a vital
constraint on its orbit through  the Local Group, under reasonable assumptions
about its proper motion and the mass distribution of the Local Group.

In estimating  the Local  Group velocity, $v_{lg}$,  of Cetus, the  first step
involves transforming  its heliocentric velocity into  a galactocentric radial
velocity, finding $v_g = -25$\,km\,s$^{-1}$.  It is assumed that the Milky Way
and M31 account for all the mass of the Local Group, and that their mass ratio
is $a  = 1$. \citet{1982MNRAS.199...67E} present arguments  which suggest that
the total angular momentum of the Local Group is close to zero, which requires
that  these two galaxies  are on  a nearly  radial orbit  with respect  to one
another.  If Cetus is  on a purely radial orbit with respect  to the centre of
mass of the Local Group then its radial velocity, $v_{lg}$, is constrained by
\begin{equation}
\mathbf{v}_{lg} \cdot \mathbf{\hat{r}} = v_g - 
\mathbf{v}_{mw} \cdot \mathbf{\hat{r}}~.
\end{equation}
where  $\mathbf{\hat{r}}$ is the  unit vector  in the  direction of  Cetus and
$\mathbf{v}_{mw}$  is the Local  Group velocity  of the  Milky Way  ($v_{mw} =
-61.5$\,km\,s$^{-1}$ for  $a =  1$).  If  it is assumed  that Cetus  is moving
radially  away   from  the   center  of  the   Local  Group  then   $v_{lg}  =
14$\,km\,s$^{-1}$, at a barycentric distance of $602$\,kpc.

\begin{figure}
\centerline{ \psfig{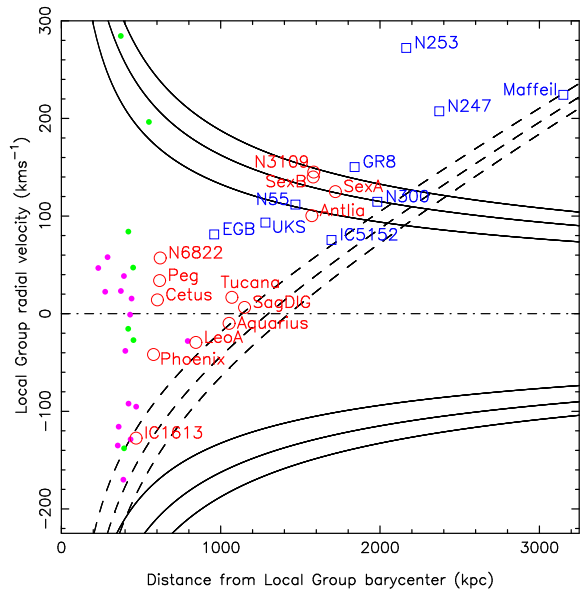}}
\caption[]{The barycentric Local Group distance verses the Local Group
radial velocity for various dwarfs in the local universe. Here, red open  
circles are  other candidate isolated  Local  Group  dwarf  galaxies,
blue  open  squares  represent galaxies belonging  to other nearby  groups,
while  magenta and green  solid circles represent  dwarf galaxies  which  
are satellites  of  M31 and  the Milky  Way, respectively.
\label{fig7}}
\end{figure}

Figure~\ref{fig7} shows the Local Group  velocity of Cetus plotted against its
barycentric  distance.  Also  shown as  red open  circles are  other candidate
isolated  Local  Group  dwarf  galaxies,  while blue  open  squares  represent
galaxies belonging  to other nearby  groups.  Magenta and green  solid circles
represent  dwarf galaxies  which  are satellites  of  M31 and  the Milky  Way,
respectively.  For  these galaxies, the approximation that  their dynamics are
governed by the net Local Group  potential is not valid, and so the velocities
derived  will not be  with respect  to the  Local Group  centre of  mass.  The
dot-dashed line represents the apocentre of  orbits in the Local Group and the
dashed curves represent  the maximum distance that a  galaxy could reach given
its current  velocity for a  total Local  Group mass of  $2, 3$ and  $4 \times
10^{12}M_\odot$  (inner to  outer  curves).  The  solid  curves represent  the
escape velocity of the Local Group  for these masses.  Thus galaxies which are
above the top set of solid lines or below the bottom set of solid lines cannot
be bound  to the Local Group.  Likewise,  galaxies to the right  of the dashed
curves would  not have  been able to  reach their  position if they  are Local
Group members~\citep{1999IAUS..192..409I}.

Several  interesting results  are immediately  apparent  in Figure~\ref{fig7};
Cetus is  close to the apocentre  of its orbit  in the Local Group,  while the
other  isolated  dSph, Tucana,  is  similarly  at  apocentre, along  with  the
Sagittarius dIrr and the transition galaxy Aquarius (DDO210).  The Sagittarius
dIrr and Aquarius  can only have reaching their current  positions if the mass
of the  Local Group is  $\gta 2 \times  10^{12}M_\odot$, and they are  on very
radial orbits. There is some overlap in this plot between Local Group galaxies
and members of nearby groups, which  reflects the well known result that their
are no  distinct boundaries between  the hundred or  so groups which  form the
Local  Supercluster.  Particularly, the  Sculptor group  joins with  the Local
Group  via a  bridge  of  galaxies including  NGC55,  UKS2323-326 and  IC5152.
Sextans A,  B, Antlia  and NGC3109 form  a distinct  grouping in the  sky (the
NGC3109 group).   Figure~\ref{fig7} shows  that they are  dynamically distinct
from other Local Group  galaxies and that they can only be  bound to the Local
Group if its mass is $\gta 4 \times 10^{12}M_\odot$.

\section{Conclusions}\label{conclusions}
This  paper has  presented a  study  of the  Cetus dwarf  galaxy, utilizing  a
matched  filter analysis  of the  INT/WFC observations  of Cetus  dwarf galaxy
presented  by  \citet{2005MNRAS.356..979M}.   The  results of  this  analysis,
however, suggests that Cetus does not possess the extra-tidal stars indicative
of any  significant interaction in the  past. Hence, it can  be concluded that
Cetus truly does represent a lonely and isolated member of the Local Group.

In  addition, this paper  has presented  a spectroscopic  survey of  the Cetus
dwarf,  utilizing a  kinematic survey  to study  its dynamical  properties.  A
maximum likelihood analysis  of the kinematic sample reveals  that Cetus has a
systemic velocity  of $-87\pm2$  ${\rm km\ s^{-1}}$  and an  internal velocity
dispersion of $17\pm2$  ${\rm km\ s^{-1}}$.  Furthermore, a  viral analysis of
this  data suggests  that this  dwarf possesses  a $M/L\sim70$  (assuming mass
traces light); however  this result may be misleading  as the data tentatively
reveals the presence  of some rotation with a best  fit rotational velocity of
$7.7\pm1.2$ ${\rm km\  s^{-1}}$; removing this rotation results  in a velocity
dispersion between 8 and 14 ${\rm km\ s^{-1}}$, dependent upon the statistical
clipping.  As ever, more data and accurate kinematic modeling will be required
to fully address  these issues. Finally, an analysis  of Cetus's orbit through
the  Local Group  find it  to  be at  apocentre, providing  a weak  additional
constraint to the mass of the Local Group. Taken in conjunction with the other
members of the  dwarf population in the local universe, the  mass of the Local
Group is found to be $\gta 2\times10^{12}M_\odot$.

\section*{Acknowledgments}
The anonymous  referee is thanked  for suggestions which improved  this paper.
GFL acknowledges the support of the Discovery Project grant DP0343508, and the
Aspen Center  for Physics,  at which  some of the  analysis presented  in this
paper was  undertaken.  Data presented herein  were obtained using  the W.  M.
Keck Observatory, which is operated as a scientific partnership among Caltech,
the University of  California and NASA.  The Observatory  was made possible by
the generous  financial support  of the W.   M.  Keck Foundation.   GFL thanks
QANTAS for his upgrade on QF8, allowing him to use his laptop effectively on a
trans-pacific flight and write a substantial proportion of this paper.

\begin{table*}
 \centering
 \begin{minipage}{140mm}
  \caption{Stars observered in this survey}
  \begin{tabular}{@{}rrrrrrrrrr@{}}
 \multicolumn{3}{c}{RA} & \multicolumn{3}{c}{Dec} & \multicolumn{1}{c}{V} &
 \multicolumn{1}{c}{I}
 & \multicolumn{1}{c}{Vel} & \multicolumn{1}{c}{$\sigma_{\rm
   Vel}$}\\ [8pt]
0 & 26 & 3.00 & -11 & 4 & 39.9 & 21.68 & 20.14 & -76.7 & 9.1 \\
0 & 26 & 9.51 & -11 & 3 & 34.3 & 21.72 & 20.24 & -77.1 & 8.7 \\
0 & 26 & 7.88 & -11 & 2 & 41.9 & 21.77 & 20.10 & -51.4 & 6.4 \\
0 & 25 & 58.13 & -11 & 5 & 14.1 & 21.90 & 20.56 & -82.6 & 6.7 \\
0 & 26 & 2.07 & -11 & 4 & 2.7 & 21.90 & 20.40 & -81.0 & 7.3 \\
0 & 26 & 8.68 & -11 & 4 & 3.7 & 21.92 & 20.41 & -84.5 & 5.8 \\
0 & 26 & 6.60 & -11 & 3 & 46.5 & 21.94 & 20.62 & -93.8 & 6.9 \\
0 & 26 & 20.46 & -11 & 4 & 17.4 & 22.05 & 20.79 & -84.8 & 14.9 \\
0 & 26 & 11.06 & -11 & 3 & 52.3 & 22.09 & 20.75 & -94.0 & 15.3 \\
0 & 26 & 1.75 & -11 & 3 & 36.4 & 22.10 & 20.83 & -83.5 & 5.3 \\
0 & 26 & 3.61 & -11 & 3 & 6.8 & 22.13 & 20.83 & -91.8 & 6.9 \\
0 & 26 & 9.69 & -11 & 2 & 9.0 & 22.15 & 20.80 & -81.0 & 9.6 \\
0 & 26 & 3.41 & -11 & 4 & 36.0 & 22.15 & 21.01 & -79.1 & 7.7 \\
0 & 25 & 58.21 & -11 & 3 & 6.4 & 22.18 & 21.08 & -73.9 & 12.6 \\
0 & 26 & 13.09 & -11 & 3 & 15.6 & 22.19 & 20.86 & -98.00 & 12.9 \\
0 & 26 & 1.21 & -11 & 2 & 41.4 & 22.20 & 20.93 & -74.7 & 6.7 \\
0 & 26 & 1.49 & -11 & 3 & 6.3 & 22.20 & 20.95 & -85.2 & 4.4 \\
0 & 26 & 32.77 & -11 & 1 & 17.8 & 22.22 & 21.00 & -99.7 & 7.5 \\
0 & 26 & 11.26 & -11 & 4 & 59.1 & 22.22 & 20.97 & -79.6 & 11.0 \\
0 & 26 & 9.20 & -11 & 2 & 10.8 & 22.22 & 20.82 & -72.7 & 11.3 \\
0 & 26 & 13.56 & -11 & 2 & 29.5 & 22.23 & 21.11 & -70.1 & 15.5 \\
0 & 25 & 59.81 & -11 & 2 & 51.4 & 22.24 & 20.90 & -89.4 & 8.7 \\
0 & 26 & 6.98 & -11 & 4 & 22.6 & 22.24 & 21.09 & -76.2 & 13.2 \\
0 & 26 & 15.75 & -11 & 3 & 34.3 & 22.24 & 21.00 & -155.7 & 16.6 \\
0 & 26 & 8.86 & -11 & 2 & 51.3 & 22.25 & 21.04 & -94.7 & 5.5 \\
0 & 26 & 10.31 & -11 & 2 & 37.4 & 22.27 & 21.00 & -61.3 & 11.9 \\
0 & 26 & 34.36 & -11 & 2 & 47.5 & 22.30 & 21.18 & -77.4 & 10.6 \\
0 & 26 & 12.57 & -11 & 3 & 14.0 & 22.33 & 21.34 & -119.2 & 1.7 \\
0 & 26 & 5.86 & -11 & 4 & 49.7 & 22.33 & 21.01 & -77.6 & 9.0 \\
0 & 26 & 15.49 & -11 & 2 & 26.4 & 22.34 & 20.97 & -83.2 & 7.7 \\
0 & 26 & 7.74 & -11 & 3 & 0.2 & 22.34 & 21.07 & -69.5 & 17.4 \\
0 & 26 & 9.35 & -11 & 2 & 25.3 & 22.34 & 21.21 & -91.0 & 11.4 \\
0 & 26 & 20.80 & -11 & 2 & 23.9 & 22.36 & 21.13 & -106.4 & 4.2 \\
0 & 26 & 6.75 & -11 & 2 & 32.6 & 22.39 & 21.18 & -78.9 & 15.7 \\
0 & 26 & 27.37 & -11 & 4 & 52.0 & 22.42 & 21.27 & -106.0 & 13.8 \\
0 & 26 & 7.08 & -11 & 2 & 32.7 & 22.44 & 21.19 & -96.9 & 5.9 \\
0 & 26 & 22.73 & -11 & 3 & 32.2 & 22.46 & 21.31 & -83.8 & 14.2 \\
0 & 25 & 58.41 & -11 & 3 & 35.1 & 22.49 & 21.33 & -74.4 & 11.0 \\
0 & 26 & 16.76 & -11 & 2 & 27.1 & 22.49 & 21.39 & -104.6 & 8.3 \\
0 & 26 & 4.41 & -11 & 3 & 4.4 & 22.50 & 21.39 & -75.4 & 11.3 \\
0 & 26 & 25.89 & -11 & 2 & 35.7 & 22.61 & 21.50 & -102.7 & 5.1 \\
0 & 26 & 7.41 & -11 & 3 & 4.9 & 22.63 & 21.43 & -71.4 & 5.0 \\
0 & 25 & 53.17 & -11 & 1 & 24.0 & 21.94 & 20.61 & -87.8 & 6.2 \\
0 & 25 & 51.36 & -11 & 3 & 5.2 & 21.98 & 20.70 & -72.8 & 7.8 \\
0 & 26 & 42.11 & -11 & 2 & 48.8 & 21.99 & 20.73 & -99.3 & 4.2 \\
0 & 25 & 56.49 & -11 & 2 & 59.5 & 22.00 & 20.80 & -78.3 & 7.4 \\
0 & 25 & 50.89 & -11 & 3 & 15.3 & 22.01 & 20.66 & -74.8 & 4.5 \\
0 & 25 & 51.93 & -11 & 4 & 42.3 & 22.04 & 20.63 & -86.7 & 11.3 \\
0 & 25 & 51.67 & -11 & 2 & 42.8 & 22.08 & 20.87 & -72.0 & 7.0 \\
0 & 25 & 56.36 & -11 & 3 & 31.1 & 22.22 & 21.05 & -62.1 & 6.9 \\
0 & 25 & 59.61 & -11 & 2 & 0.8 & 22.29 & 21.08 & -68.9 & 6.4 \\
\end{tabular}
\end{minipage}
\end{table*}

\begin{table*}
 \centering
 \begin{minipage}{140mm}
  \caption{Stars observered in this survey}
  \begin{tabular}{@{}rrrrrrrrrr@{}}
0 & 25 & 55.02 & -11 & 3 & 49.5 & 22.32 & 21.18 & -175.4 & 3.8 \\
0 & 26 & 48.57 & -11 & 5 & 3.4 & 22.37 & 21.19 & 120.3 & 8.7 \\
0 & 25 & 55.47 & -11 & 3 & 27.1 & 22.59 & 21.62 & 183.7 & 9.2 \\
0 & 26 & 3.19 & -11 & 4 & 31.2 & 22.68 & 21.57 & -211.1 & 9.9 \\
0 & 26 & 13.74 & -11 & 3 & 14.1 & 22.68 & 21.61 & 227.5 & 3.6 \\
0 & 26 & 2.18 & -11 & 2 & 38.1 & 22.69 & 21.66 & -56.1 & 7.2 \\
0 & 25 & 57.69 & -11 & 3 & 5.7 & 22.71 & 21.58 & -88.7 & 9.6 \\
0 & 26 & 18.48 & -11 & 3 & 19.0 & 22.76 & 21.62 & -123.3 & 5.7 \\
0 & 25 & 56.72 & -11 & 2 & 36.0 & 22.78 & 21.55 & 172.5 & 7.4 \\
0 & 25 & 53.76 & -11 & 1 & 46.8 & 22.79 & 21.70 & -91.5 & 15.3 \\
0 & 26 & 50.97 & -11 & 3 & 47.3 & 22.80 & 21.62 & 117.4 & 8.6 \\
0 & 26 & 26.24 & -11 & 2 & 7.3 & 22.89 & 21.70 & -104.9 & 8.8 \\
0 & 25 & 54.15 & -11 & 5 & 35.4 & 22.89 & 21.97 & 189.6 & 4.6 \\
0 & 26 & 8.15 & -11 & 3 & 7.7 & 22.91 & 21.85 & -71.4 & 6.3\\
0 & 25 & 46.41 & -11 & 3 & 51.8 & 22.94 & 21.84 & 253.5 & 12.9 \\
0 & 25 & 59.34 & -11 & 2 & 5.6 & 22.96 & 21.99 & -119.4 & 17.9 \\
0 & 25 & 54.71 & -11 & 3 & 0.5 & 22.99 & 21.94 & -129.5 & 2.7 \\
0 & 26 & 40.69 & -11 & 2 & 24.6 & 23.00 & 21.83 & -51.6 & 7.2 \\
0 & 25 & 57.79 & -11 & 1 & 13.9 & 23.09 & 21.61 & 2.6 & 6.4 \\
\end{tabular}
\end{minipage}
\end{table*}


\begin{thebibliography}{DEM}
\bibitem[\protect\citeauthoryear{Belokurov et 
al.}{2006}]{2006ApJ...642L.137B} Belokurov V., et al., 2006, ApJ, 642, L137 

\bibitem[\protect\citeauthoryear{Belokurov et 
al.}{2006}]{2006ApJ...647L.111B} Belokurov V., et al., 2006, ApJ, 647, L111 

\bibitem[\protect\citeauthoryear{Bouchard, Carignan, \& 
Staveley-Smith}{2006}]{2006AJ....131.2913B} Bouchard A., Carignan C., 
Staveley-Smith L., 2006, AJ, 131, 2913 

\bibitem[\protect\citeauthoryear{Einasto, Kaasik, \& 
Saar}{1974}]{1974Natur.250..309E} Einasto J., Kaasik A., Saar E., 1974, 
Natur, 250, 309 

\bibitem[\protect\citeauthoryear{Einasto \& 
Lynden-Bell}{1982}]{1982MNRAS.199...67E} Einasto J., Lynden-Bell D., 1982, 
MNRAS, 199, 67 

\bibitem[\protect\citeauthoryear{Faber et al.}{2003}]{2003SPIE.4841.1657F} 
Faber S.~M., et al., 2003, SPIE, 4841, 1657 

\bibitem[\protect\citeauthoryear{Freeman \& 
Bland-Hawthorn}{2002}]{2002ARA&A..40..487F} Freeman K., Bland-Hawthorn J., 
2002, ARA\&A, 40, 487 

\bibitem[\protect\citeauthoryear{Grillmair \& 
Dionatos}{2006}]{2006ApJ...641L..37G} Grillmair C.~J., Dionatos O., 2006, 
ApJ, 641, L37 

\bibitem[\protect\citeauthoryear{Harbeck, Gallagher, \& 
Grebel}{2004}]{2004AJ....127.2711H} Harbeck D., Gallagher J.~S., III, 
Grebel E.~K., 2004, AJ, 127, 2711 

\bibitem[\protect\citeauthoryear{Ibata, Gilmore, \& 
Irwin}{1994}]{1994Natur.370..194I} Ibata R.~A., Gilmore G., Irwin M.~J., 
1994, Natur, 370, 194  

\bibitem[\protect\citeauthoryear{Ibata et al.}{2001}]{2001Natur.412...49I} 
Ibata R., Irwin M., Lewis G., Ferguson A.~M.~N., Tanvir N., 2001, Natur, 
412, 49  

\bibitem[\protect\citeauthoryear{Ibata et al.}{2005}]{2005ApJ...634..287I} 
Ibata R., Chapman S., Ferguson A.~M.~N., Lewis G., Irwin M., Tanvir N., 
2005, ApJ, 634, 287 

\bibitem[\protect\citeauthoryear{Irwin}{1999}]{1999IAUS..192..409I} Irwin 
M., 1999, IAUS, 192, 409 

\bibitem[\protect\citeauthoryear{Johnston, Choi, \& 
Guhathakurta}{2002}]{2002AJ....124..127J} Johnston K.~V., Choi P.~I., 
Guhathakurta P., 2002, AJ, 124, 127 

\bibitem[\protect\citeauthoryear{Kepner et al.}{1999}]{1999ApJ...517...78K} 
Kepner J., Fan X., Bahcall N., Gunn J., Lupton R., Xu G., 1999, ApJ, 517, 
78 

\bibitem[\protect\citeauthoryear{Klypin et al.}{1999}]{1999ApJ...522...82K} 
Klypin A., Kravtsov A.~V., Valenzuela O., Prada F., 1999, ApJ, 522, 82  

\bibitem[\protect\citeauthoryear{Martin et al.}{2004}]{2004MNRAS.348...12M} 
Martin N.~F., Ibata R.~A., Bellazzini M., Irwin M.~J., Lewis G.~F., Dehnen 
W., 2004, MNRAS, 348, 12 

\bibitem[\protect\citeauthoryear{Mateo}{1998}]{1998ARA&A..36..435M} Mateo 
M.~L., 1998, ARA\&A, 36, 435 

\bibitem[\protect\citeauthoryear{Mayer et al.}{2001a}]{2001ApJ...559..754M} 
Mayer L., Governato F., Colpi M., Moore B., Quinn T., Wadsley J., Stadel 
J., Lake G., 2001a, ApJ, 559, 754 


\bibitem[\protect\citeauthoryear{Mayer et al.}{2001b}]{2001ApJ...547L.123M} 
Mayer L., Governato F., Colpi M., Moore B., Quinn T., Wadsley J., Stadel 
J., Lake G., 2001b, ApJ, 547, L123 

\bibitem[\protect\citeauthoryear{Mayer et al.}{2002}]{2002MNRAS.336..119M} 
Mayer L., Moore B., Quinn T., Governato F., Stadel J., 2002, MNRAS, 336, 
119 

\bibitem[\protect\citeauthoryear{Mayer et al.}{2006}]{2006MNRAS.369.1021M} 
Mayer L., Mastropietro C., Wadsley J., Stadel J., Moore B., 2006, MNRAS, 
369, 1021 

\bibitem[\protect\citeauthoryear{McConnachie et 
al.}{2005}]{2005MNRAS.356..979M} McConnachie A.~W., Irwin M.~J., Ferguson 
A.~M.~N., Ibata R.~A., Lewis G.~F., Tanvir N., 2005, MNRAS, 356, 979 

\bibitem[\protect\citeauthoryear{McConnachie \& 
Irwin}{2006}]{alaninprep} McConnachie A.~W., Irwin M.~J., 2006, 
MNRAS, 365, 1263 

\bibitem[\protect\citeauthoryear{McConnachie, Arimoto, Irwin \& 
Tolstoy}{2006}]{alaninprep2} McConnachie A.~W., Arimoto, N., Irwin M.~J., Tolstoy, E., 2006, 
MNRAS, {In press}, astro-ph/0609237

\bibitem[\protect\citeauthoryear{Moore et al.}{1999}]{1999ApJ...524L..19M} 
Moore B., Ghigna S., Governato F., Lake G., Quinn T., Stadel J., Tozzi P., 
1999, ApJ, 524, L19 

\bibitem[\protect\citeauthoryear{Richstone \& 
Tremaine}{1986}]{1986AJ.....92...72R} Richstone D.~O., Tremaine S., 1986, 
AJ, 92, 72 

\bibitem[\protect\citeauthoryear{Robin et al.}{2003}]{2003A&A...409..523R} 
Robin A.~C., Reyl{\'e} C., Derri{\`e}re S., Picaud S., 2003, A\&A, 409, 523 

\bibitem[\protect\citeauthoryear{Rockosi et 
al.}{2002}]{2002AJ....124..349R} Rockosi C.~M., et al., 2002, AJ, 124, 349 

\bibitem[\protect\citeauthoryear{Rutledge, Hesser, \& 
Stetson}{1997}]{1997PASP..109..907R} Rutledge G.~A., Hesser J.~E., Stetson 
P.~B., 1997, PASP, 109, 907 

\bibitem[\protect\citeauthoryear{Sarajedini et 
al.}{2002}]{2002ApJ...567..915S} Sarajedini A., et al., 2002, ApJ, 567, 915  
 
\bibitem[\protect\citeauthoryear{Stoehr et al.}{2002}]{2002MNRAS.335L..84S} 
Stoehr F., White S.~D.~M., Tormen G., Springel V., 2002, MNRAS, 335, L84 

\bibitem[\protect\citeauthoryear{Venn et al.}{2004}]{2004AJ....128.1177V} 
Venn K.~A., Irwin M., Shetrone M.~D., Tout C.~A., Hill V., Tolstoy E., 
2004, AJ, 128, 1177 

\bibitem[\protect\citeauthoryear{White \& Rees}{1978}]{1978MNRAS.183..341W} 
White S.~D.~M., Rees M.~J., 1978, MNRAS, 183, 341 

\bibitem[\protect\citeauthoryear{Whiting, Hau, \& 
Irwin}{1999}]{1999AJ....118.2767W} Whiting A.~B., Hau G.~K.~T., Irwin M., 
1999, AJ, 118, 2767  

\bibitem[\protect\citeauthoryear{Willman et 
al.}{2005a}]{2005ApJ...626L..85W} Willman B., et al., 2005a, ApJ, 626, L85 


\bibitem[\protect\citeauthoryear{Willman et 
al.}{2005b}]{2005AJ....129.2692W} Willman B., et al., 2005b, AJ, 129, 2692 

\bibitem[\protect\citeauthoryear{Zucker et al.}{2004}]{2004ApJ...612L.121Z} 
Zucker D.~B., et al., 2004, ApJ, 612, L121 

\bibitem[\protect\citeauthoryear{Zucker et al.}{2006}]{2006ApJ...643L.103Z} 
Zucker D.~B., et al., 2006, ApJ, 643, L103 

\bibitem[\protect\citeauthoryear{Zucker et al.}{2006}]{2006ApJ...650L..41Z} 
Zucker D.~B., et al., 2006, ApJ, 650, L41 
 
\end{thebibliography}
\end{document}